\documentclass[aps,prd,reprint,superscriptaddress,showkeys,amsmath,amssymb]{revtex4-2}

\usepackage{graphicx}

\providecommand{\vv}[1]{\vec{#1}}

\usepackage{hyperref}
\begin{document}

\title{GW Microlensing: Degeneracy with Unlensed Precessing and Non-Spinning Gravitational-Wave Signals}

\author{Disha Hegde}
\email{disha.hegde@uclouvain.be}
\affiliation{University of Louvain, Centre for Cosmology, Particle Physics and Phenomenology-CP3, Chemin du Cyclotron 2, 1348 Louvain-la-Neuve, Belgium}
\affiliation{Sardar Vallabhbhai National Institute of Technology, Surat, Gujarat 395007, India}

\author{Anupreeta More}
\affiliation{The Inter-University Centre for Astronomy and Astrophysics, Post Bag 4, Ganeshkhind, Pune 411007, India}
\affiliation{Kavli Institute for the Physics and Mathematics of the Universe (IPMU), 5-1-5 Kashiwanoha, Kashiwa-shi, Chiba 277-8583, Japan}

\begin{abstract}
With nearly 400 Gravitational Wave (GW) events detected by the LIGO--Virgo--Kagra detector network and many more expected in the future, morphological similarities between GWs produced by different astrophysical effects and/or source parameters could pose challenges for template-based searches and subsequent model parameter inferences. Specifically, the modulation in a GW signal resulting from the precession of a compact binary system may resemble the beating pattern expected in microlensed GW signals. The latter is gravitational lensing effect arising from a  massive ($10-10^5$~M$_\odot$) compact object acting as a deflector relevant to the LIGO frequency band. We investigate any degeneracies in the observational imprints arising from these two effects and attempt to distinguish them with the help of machine learning-based classification techniques.
To this end, we generate a sample of 20,000 GW signals of each class with a network optimal SNR above a threshold of 20. We use a Convolutional Neural Network (CNN) architecture to train on time-frequency spectrograms (Q-Transform images) and achieve a performance: up to 95\% accuracy in Gaussian noise and 82\% in real noise. We further explored classification between microlensed (ML) and unlensed non-spinning (UN) signals since the latter kind is detected routinely in the GW data. For completeness, we also attempted to classify unlensed non-spinning (UN) vs. unlensed precessing (UP). Distinguishing UN from UP proves challenging, even in Gaussian noise while classification between ML and UN reaches up to 80\% accuracy in real noise.
We identified regions of parameter space where the model performs well. Finally, we evaluated our ML vs. UN network on real GW events, finding reasonable performance for the model trained on Gaussian noise compared to real noise. This work presents the first pipeline for identifying microlensed signals in the GW data at low latency.
\end{abstract}

\maketitle

\section{Introduction}
Gravitational waves (GWs), often described as ripples in the fabric of space-time, are predictions of Einstein's General Theory of Relativity (GR).
The accelerated motion of any massive object leads to disturbances in the geometry of space-time, which propagate away from the source with the speed of light and are known as gravitational waves. The strongest sources of GWs are known to be the collisions or mergers of compact binary objects \citep[]{buonanno2015sourcesgravitationalwavestheory}, like black holes (BHs) and neutron stars (NS).  The first direct detection of GWs took place on September 14, 2015 with the Laser Interferometer Gravitational-wave Observatory (LIGO), which measured the fractional change in lengths of its two arms (strain) caused by the generation of GWs by a binary black hole (BBH) merger \citep[]{PhysRevLett.116.061102}. Since then, the LIGO-Virgo-Kagra (LVK) network of detectors has completed three observing runs, detecting 90 confident GW events \citep[]{PhysRevX.13.041039} and improved constraints on the mass distributions and merger rate density models of the binary population \citep[]{PhysRevX.13.011048}. The ongoing fourth observing run (O4) has substantially expanded this catalog, with the number of candidate GW events now approaching 400 \citep[]{2025arXiv250818082T, 2026arXiv260527225T}.

The morphology of a GW signal depends on a number of intrinsic source properties like masses, spins, and eccentricity, as well as extrinsic properties like sky location, detector orientation, and distance to the source.
When the spin of, at least, one of the component masses in the binary, is misaligned with the orbital angular momentum vector ($\vv{L}$), a relativistic interaction between the spin and $\vv{L}$ induces a rotation of the orbital plane around a fixed axis, called precession. As a result, the emission of GWs is maximized along the direction of the orbital angular momentum ($\hat{L}$). However, since $\hat{L}$ evolves over time, an inertial observer detects a time-dependent modulation of the GW amplitude and phase as the orbital plane alternates between pointing towards and away from the observer \citep[]{PhysRevD.49.6274}. This modulation, characteristic of precession, is observable with current GW detectors \citep[]{PhysRevD.110.023038}. Comprehensive discussions of spin-precession dynamics and its imprint on gravitational-wave signals can be found in \citep{2003PhRvD..67j4025B}. Precession has also been constrained in several LIGO--Virgo observations, most notably in GW190412 and GW200129, where in-plane spin components produce measurable waveform modulations \citep{PhysRevD.102.043015, Hannam_2022}.

Astrophysical effects, external to the binary, like gravitational lensing can also influence the observed GW signals.
In gravitational lensing, the GW signals encounter massive objects on their way to the observer such that the signals can be redirected and focused, leading to a magnification of their amplitudes and slowed down differentially, introducing time delays between multiple paths taken by the GW signals.
For GWs detected within the LIGO frequency range of 10~Hz$-10^4$~Hz, the geometric optics approximation proves effective when the lensing object is a galaxy or galaxy cluster.
However, for lens masses falling within the range of 10~M$_{\odot}-10^5$~M$_{\odot}$, referred to as microlensing, the wavelength of GWs approaches the Schwarzschild radius of the lens mass. This leads to significant wave optics effects, resulting in frequency-dependent modulations in the GW signals \citep[e.g.,][]{meena2019,Meena_2022}. These microlenses, low mass compact objects, could be located in our Milky Way or somewhere in the line-of-sight to the source either as free-floating objects in isolated environments or embedded in another intervening galaxy.

Analysis and interpretation of GW signals with frequency-dependent modulations can thus be complicated because completely different physical effects like microlensing and spin precession may mimic each other, especially, in the presence of noise as seen in the real data.
In fact, a related study by \citet{Liu_2024} investigated similarities between GW signals from unlensed precessing binaries and those affected by microlensing ($10^2$--$10^5,M_\odot$) although geometric optics approximation was applied and only a few illustrative cases were analysed.
Such degeneracies, if any, can challenge template-based searches and parameter estimation by leading to misidentification of the underlying signal morphology and biased astrophysical inference \citep{Chan_2025}.
In this work, we generate the microlensed non-spinning GWs by including the accurate modulations arising from the wave optics effect and unlensed precessing binaries using astrophysically realistic statistical populations for comparison and classification.

Systematic searches for microlensing in the LIGO--Virgo data have been carried out and the non-detection has been used to place constraints on the dark matter \citep[e.g.][]{2021ApJ...923...14A, 2024ApJ...970..191A, 2025arXiv251216347T, 2023MNRAS.526.3832J, 2022ApJ...926L..28B,2025ApJ...990...68C}. The algorithms used in these searches follow Bayesian analysis to perform full-scale parameter estimation to identify microlensing features \citep[e.g.][]{wright2022gravelamps}, although alternative model-independent search strategies have recently been proposed \citep{2025ApJ...984..107C}. Such methodologies can be time consuming and are not scalable with increasing sample sizes of GW population as expected from improved detector sensitivites in the near future. The same goes for searching for precessing binaries \citep[e.g.][]{2017PhRvD..95f4056I, 2023PhRvD.108l3016M, PhysRevD.110.023038}. Thus, faster algorithms are needed to sift through a large sample to identify promising GW event candidates that are likely to show microlensing and discriminate them from either unlensed precessing or non-spinning signals (the latter being the more typically detected population).

In the past few years, the use of machine learning for astrophysical research has rapidly increased, given its effectiveness in pattern recognition and handling `big data'. Consequently, machine learning techniques have been widely employed in GW astronomy for applications like searches, noise detection and classification, parameter estimation, and waveform modelling (see reviews e.g., \cite{Cuoco_2020,cuoco2024}. Convolutional neural networks (CNNs) have become the norm for dealing with computer vision problems, due to their ability to extract and learn complex features from image data with great accuracy. Several studies have demonstrated the potential of CNNs for classification tasks in GW astronomy \citep[e.g.][]{PhysRevD.104.124057, PhysRevD.107.024030, Magare_2024,8cz1-kl6n}, achieving performance comparable to the respective traditional pipelines but with significantly reduced computational cost.
Thus, we decide to use machine learning networks in this work.

In terms of the type of binary population, we restrict ourselves to BBHs as they can be detected at large distances and are more likely to be microlensed than the binaries comprising of a NS. We attempt to develop a deep learning algorithm that can help distinguish between microlensing and precession. The rest of the paper is structured as follows. Section~\ref{sec:method} describes the data generation methods and the deep learning model architecture used. Section~\ref{sec:res} reports the results of the analyses carried out and outlines the inferences made. Finally, Section~\ref{sec:sum} summarizes the findings and conclusions drawn from this work.

\section{Methodology}
\label{sec:method}
In this section, we describe the design of the datasets and the procedure followed for generating them. We also describe the neural network architecture considered in this work as well as the specifics of the training process.

\subsection{Data Generation}
\label{sec:datagen}

A mock dataset of GW signals was generated using \textsc{GWMAT}\footnote{\href{https://git.ligo.org/anuj.mishra/gwmat/}{https://git.ligo.org/anuj.mishra/gwmat/}}, comprising a total of 60,000 events categorized into 20,000 unlensed non-spinning (UN), 20,000 unlensed precessing (UP), and 20,000 microlensed non-spinning (ML) BBH signals. BBH parameters were derived from a population model constructed based on the GWTC-3 catalog \citep[]{PhysRevX.13.041039}. The \textsc{IMRPhenomXPHM} \citep[]{PhysRevD.103.104056} approximant was used to model the signals and an optimal network SNR threshold of 20 was applied. For ML signals an additional condition was applied on the time delay,  $t_d < 0.15s$, to avoid scenarios where the lensing-induced waveform overlap becomes negligible, ensuring that microlensing effects remain observable in the signal morphology.

A statistical fit to the distributions of key observed parameters, including component masses, spin magnitudes, spin tilt angles, and the redshift distribution of BBH mergers, was provided by the population model. All other BBH parameters were sampled uniformly from their respective domains. For microlensing parameters, a log-uniform prior for the lens mass (in units of $M_{\odot}$) and a power-law prior with index 1 for the impact parameter were adopted, motivated by geometrical considerations and isotropy \citep[]{PhysRevD.98.083005}. The source merger rate density was modelled using the Madau--Dickinson profile \citep[]{annurev:/content/journals/10.1146/annurev-astro-081811-125615}, and a source redshift distribution as described by \citep[]{2018ApJ...863L..41F} was assumed. The \textsc{Bilby} \citep[]{bilby_paper} and \textsc{Gwpopulation} \citep[]{2019PhRvD.100d3030T} packages were used to generate these population models.

The source redshift range was set to $z$ $\in (0.001,1.71)$ for UN/UP signals, and  $z$ $\in  (0.001,10)$ for ML signals. The lower limit corresponds to a value below which merger rate is negligible due to limited cosmological volume and low star formation rate. The upper limits of 1.71 and 10 for UN/UP and ML signals, respectively, were used to approximate the maximum detection distances achievable with current ground-based detectors under ideal conditions, such as low impact parameters, high lens mass, and massive BBH systems. Detailed parameter distributions and sampling ranges are provided in  Table~\ref{tab:1}.

\begin{table}[b]
\caption{Parameter prior distributions and sampling ranges adopted for the simulated gravitational-wave signals.}
\label{tab:1}
\begin{ruledtabular}
\begin{tabular}{lll}
\textbf{Parameter} & \textbf{Range} & \textbf{Prior distribution} \\
\hline
Mass $(M_\odot)$ & $(4.98,\,112.5)$ & Single-peak smoothed \\
Spin tilt $(\mathrm{rad})$ & $(0,\,\pi)$ & Gaussian isotropic \\
Spin magnitude & $(0,\,1)$ & Beta \\
Lens mass $(M_\odot)$ & $(10,\,10^5)$ & Log-uniform \\
Impact parameter $y$ & $(0.01,\,1)$ & Power law (index $=1$) \\
$\theta_{\rm jn}$ $(\mathrm{rad})$ & $(0,\,\pi)$ & Sine \\
Declination $(\mathrm{rad})$ & $(-\pi/2,\,\pi/2)$ & Cosine \\
Right ascension $(\mathrm{rad})$ & $(0,\,2\pi)$ & Uniform \\
$\phi$ $(\mathrm{rad})$ & $(0,\,2\pi)$ & Uniform (periodic) \\
Phase $(\mathrm{rad})$ & $(0,\,2\pi)$ & Uniform (periodic) \\
Polarization $(\mathrm{rad})$ & $(0,\,2\pi)$ & Uniform (periodic) \\
Redshift (UN/UP) & $(10^{-3},\,1.71)$ & Fishbach (2018) \\
Redshift (ML) & $(10^{-3},\,10)$ & Madau--Dickinson SFR \\
\end{tabular}
\end{ruledtabular}
\end{table}

The waveforms generated using these realistic parameter distributions, were then projected onto the antenna pattern of the H1 detector. Two datasets were subsequently constructed: one with the signals buried in simulated Gaussian noise, and another with the signals injected into real detector noise. For the Gaussian noise dataset, the detector noise power spectral density (PSD) corresponding to the O4 sensitivity curve of H1 \citep[]{2016LRR....19....1A} was used.
For the real-noise dataset, $\sim 32,000$~sec of H1 data from the O4a observing run were used, from which 8-~sec segments were extracted with a maximum overlap of 4~sec between consecutive segments, and the signals were injected into these segments. It was made sure that the signal peak occurs at t=~sec for both the cases.

Q-Transforms were generated over a time window of $(-0.5, 0.15)$ seconds around the signal peak. For signals injected into Gaussian noise, the settings used were: $q$-range = (4, 64), \texttt{whiten=False}, and \texttt{norm=`mean'}. For signals injected into real detector noise, the parameters were: $q$-range = (4, 15), \texttt{whiten=True}, and \texttt{highpass=True}, with all other settings kept at their default values. These configurations were selected to optimize the visual clarity of the signal features in each case. Specifically, the broader $q$-range and lack of whitening were suitable for the cleaner, stationary background of Gaussian noise, while the narrower $q$-range and whitening were necessary to suppress non-stationary noise artifacts in real data and enhance the visibility of the injected signal.

\begin{figure}
        \includegraphics[width=\columnwidth]{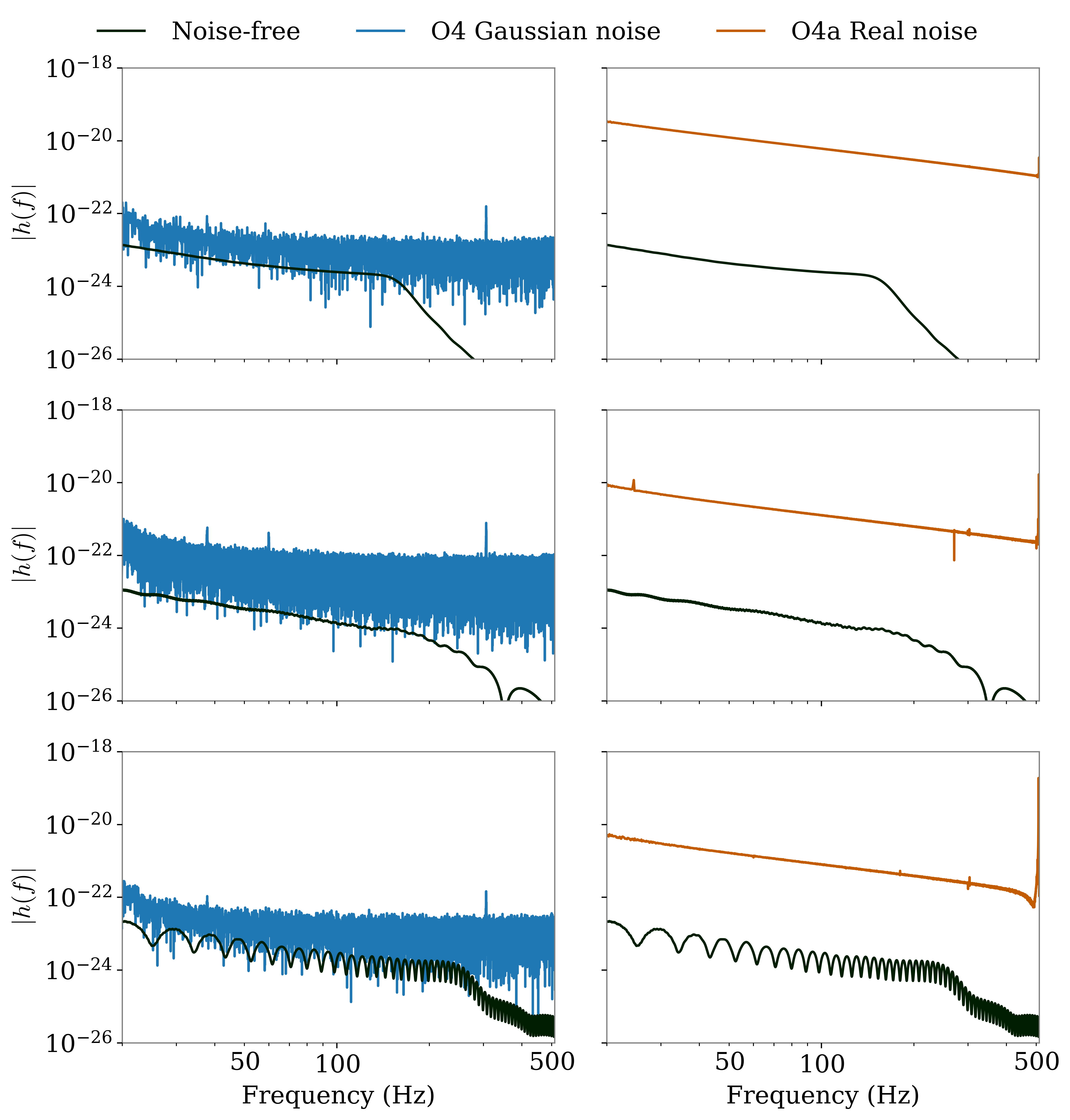}
        \caption{Mock frequency series for simulated gravitational-wave signals injected into O4 Gaussian noise (left column) and real O4a noise (right column), overlaid with the corresponding noise-free signals (in black) for reference, for unlensed non-spinning (top row), unlensed precessing (middle row), and microlensed non-spinning (bottom row) signals.}
        \label{fig:2.1}
\end{figure}

\begin{figure}
        \includegraphics[width=\columnwidth]{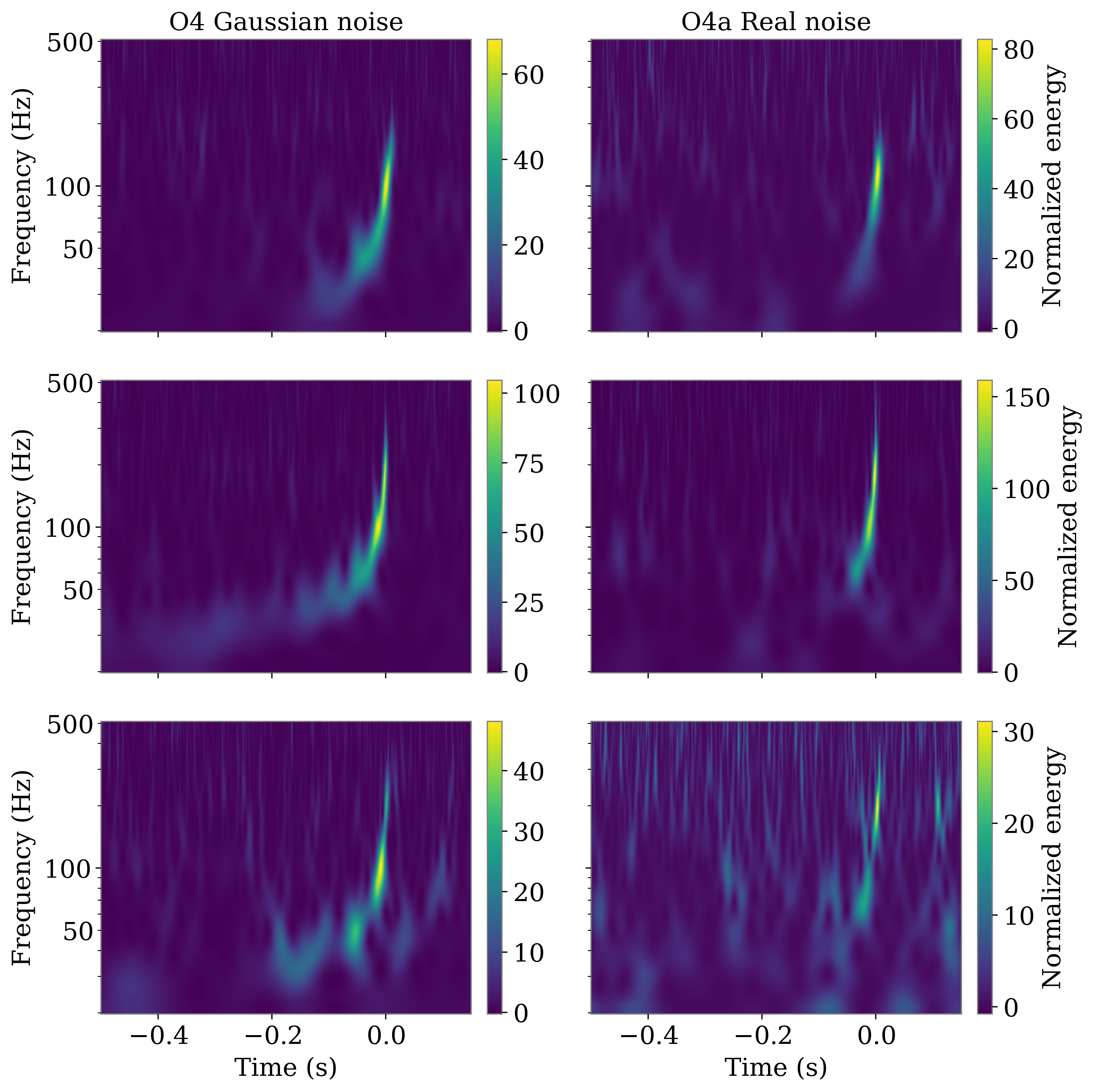}
        \caption{Example Q-transforms for simulated gravitational-wave signals injected into O4 Gaussian noise (left column) and real O4a noise (right column) for unlensed non-spinning (top row), unlensed precessing (middle row), and microlensed non-spinning (bottom row) signals.}
        \label{fig:2.2}
\end{figure}

Finally, one example each of the UN, UP, and ML signals is presented in two formats: frequency series (FS) and Q-Transforms (QTs) in Figures~\ref{fig:2.1} and ~\ref{fig:2.2}, respectively. The frequency series are shown for three cases---noise-free, signals in O4 Gaussian noise, and signals in O4a real noise---while the Q-Transforms are displayed for two cases: O4 Gaussian noise and  O4a real noise.
The frequency range is restricted to $(20$ -- $512)$ Hz.
The binary and lensing parameters corresponding to these representative signals are summarised in Table~\ref{tab:example_params}. During training, the strain amplitude is limited to the range $10^{-26}$ -  $10^{-20}$ for signals injected into O4 Gaussian noise, and $10^{-24}$ - $10^{-20}$ for those injected into O4a real detector noise.

\begin{table}[b]
\caption{Parameters of the example unlensed nonprecessing (UN), unlensed precessing (UP), and microlensed (ML) signals shown in Figs.~\ref{fig:2.1} and \ref{fig:2.2}.}
\label{tab:example_params}
\begin{ruledtabular}
\begin{tabular}{lccc}
\textbf{Parameter} & \textbf{UN} & \textbf{UP} & \textbf{ML} \\
\hline
$m_1\,(M_\odot)$ & 61.27 & 41.67 & 35.00 \\
$m_2\,(M_\odot)$ & 54.01 & 22.70 & 25.93 \\
$a_1$            & --    & 0.07  & -- \\
$a_2$            & --    & 0.32  & -- \\
$\theta_1\,(\mathrm{deg})$ & -- & 65.41 & -- \\
$\theta_2\,(\mathrm{deg})$ & -- & 67.83 & -- \\
$\rho$           & 34.72 & 52.20 & 33.39 \\
$m_l\,(M_\odot)$ & --    & --    & 4746.82 \\
$y_l$            & --    & --    & 0.58 \\
$t_d\,(\mathrm{s})$ & -- & -- & 0.11 \\
\end{tabular}
\end{ruledtabular}
\end{table}

\subsection{Neural Network Architecture and Training}
\label{sec:2.2}
The CNN architectures used for training on frequency series and Q-Transform images are illustrated in Figures ~\ref{fig:2.1} and ~\ref{fig:2.2}, respectively. Both models are adapted from the 6-block CNN proposed in SLICK by \citet{Magare_2024}, which was designed to distinguish strongly lensed GW signals from unlensed ones using sine-Gaussian projection maps.

\begin{figure}[htbp]
        \includegraphics[width=\columnwidth]{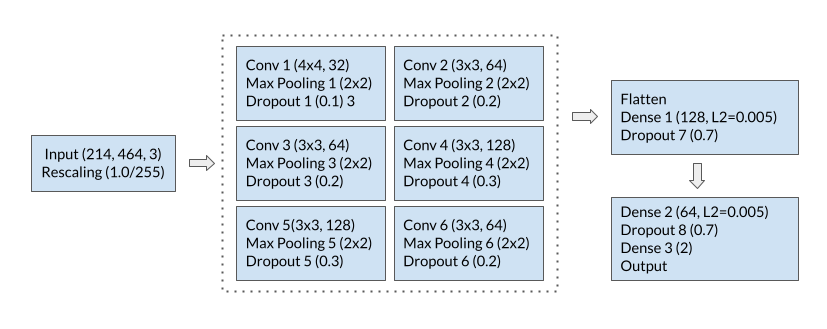}
        \caption{The architecture of the 6-block CNN used for classification of signals input as QT images}
        \label{fig:2.4}
\end{figure}

The first architecture (CNN-1), employed for FS images, consists of five convolutional blocks followed by a flattening layer, two fully connected (dense) blocks, and an output layer. Each convolutional block comprises a convolutional layer with ReLU activation, a max-pooling layer to reduce spatial dimensionality, and a dropout layer to reduce overfitting. All convolutional layers use a kernel size of $(3 \times 3)$, except for the first layer, which uses a $(4 \times 4)$ kernel. The max-pooling layers use a kernel size of $(2 \times 2)$. The fully connected blocks include a dense layer with L2 regularization succeeded by a dropout layer. The final output layer contains two units with a softmax activation, enabling binary classification.
The second architecture (CNN-2), used for QT images shown in Figure ~\ref{fig:2.4} , is identical to CNN-1 in structure but includes an additional convolutional block with 64 filters before the flattening layer. Both the models have a rescaling layer in the beginning to ensure that input pixel values are normalized to the [0, 1] range, which helps stabilize training and improves convergence.

While the primary objective was to classify ML and UP signals using images of FS and QTs, we also explored the classification of UN versus UP signals, as well as ML versus UN signals using QT images. For all experiments, we adopt a consistent binary labeling convention in which class 1 corresponds to unlensed (UN) signals and class 0 corresponds to microlensed (ML) signals.
The CNN outputs a single probability, interpreted as $p(\text{UN})$ for ML vs. UN classification and $p(\text{UP})$ for ML vs. UP classification, depending on the task. A classification threshold $t$ is applied on the corresponding probability, such that events with $p>t$ are assigned to class 1 (UN or UP), and otherwise to class 0 (ML). The dataset consisted of 12,000 images per class for training, with 4,000 images per class reserved for validation and another 4,000 for testing. The networks were trained using the binary cross-entropy loss function, which is well-suited for binary classification tasks. Optimization was performed using the Adam optimizer \citep[]{kingma2017adammethodstochasticoptimization} with the default learning rate of 0.001. Model performance was evaluated using the `accuracy' metric. The training and evaluation processes were carried out using \textsc{keras} \citep[]{chollet2015keras} and \textsc{scikit-learn} \citep[]{scikit-learn}.

CNN-1 was used to train on FS images of ML and UP signals under three conditions: noise-free, Gaussian noise-injected, and real noise-injected. The input images were $114 \times 564$ pixel grayscale images, fed in batches of 1,000, and trained for 100 epochs with early stopping (patience of 10). The model with the lowest validation loss was saved. CNN-2 was used for all QT-based experiments, including the classification of  ML vs. UP, ML vs. UN, and UN vs. UP signals, under two noise conditions: Gaussian and real noise. The input images were $214 \times 464$ pixel RGB images, processed in batches of 300, and trained for 100 epochs with early stopping (patience of 10). As with CNN-1, the model with the lowest validation loss was retained.

\section{Results}
\label{sec:res}
We report the performance and evaluation results of the trained models for three experiments: ML vs. UP, ML vs. UN, and UN vs. UP.  In each case, the positive class corresponds to the second class listed (e.g., UP in ML vs. UP). Accuracy and loss values for both the training and validation datasets were tracked at each epoch to monitor model performance over time. To further evaluate the model's classification capabilities, Receiver Operating Characteristic (ROC) curves were generated. These curves plot the true positive rate (TPR) against the false positive rate (FPR) across different detection thresholds, showing the model's ability to balance these two quantities. A higher area under the curve (AUC) indicates stronger overall performance, as the model more effectively distinguishes between classes. Additionally, predictions on the test data were used to construct confusion matrices. These matrices provide a detailed breakdown of the model's predictions, showing true positives, false positives, true negatives, and false negatives. This allows for the identification of where the model excels and where it may misclassify.

The methodology developed in this work has also been implemented into a dedicated pipeline, \textsc{RaSMUNN} (Rapid Search for Microlensing Using Neural Networks), which enables the application of the trained ML vs. UN classifier directly to real gravitational-wave data. The primary goal of this pipeline is to provide a low-latency classification framework that can rapidly identify and discard signals that are clearly consistent with the unlensed (UN) hypothesis, allowing computational resources to be focused on more promising candidates for detailed microlensing analyses.

\subsection{Microlensed non-spinning vs. Unlensed Precessing}
CNN-1 was trained on frequency series (FS) images under three conditions: noise-free, Gaussian noise, and real noise. On noise-free FS, the model performed exceptionally well, achieving over 99\% accuracy and an AUC of 1, correctly classifying nearly all signals. However, when trained on FS injected with Gaussian or real noise, the model failed to learn meaningful patterns---training and validation accuracies plateaued around 50\%, and loss values quickly stabilized at $\sim$ 0.69, indicating random guessing. Although different architectures and training strategies were tested, the results remained unchanged. This consistent under-performance across noisy datasets suggested that FS image representations may be inadequate for distinguishing classes with the architectures we explored, motivating a shift towards a different representation: the Q-Transform (QT).

\begin{figure}
        \includegraphics[width=\columnwidth]{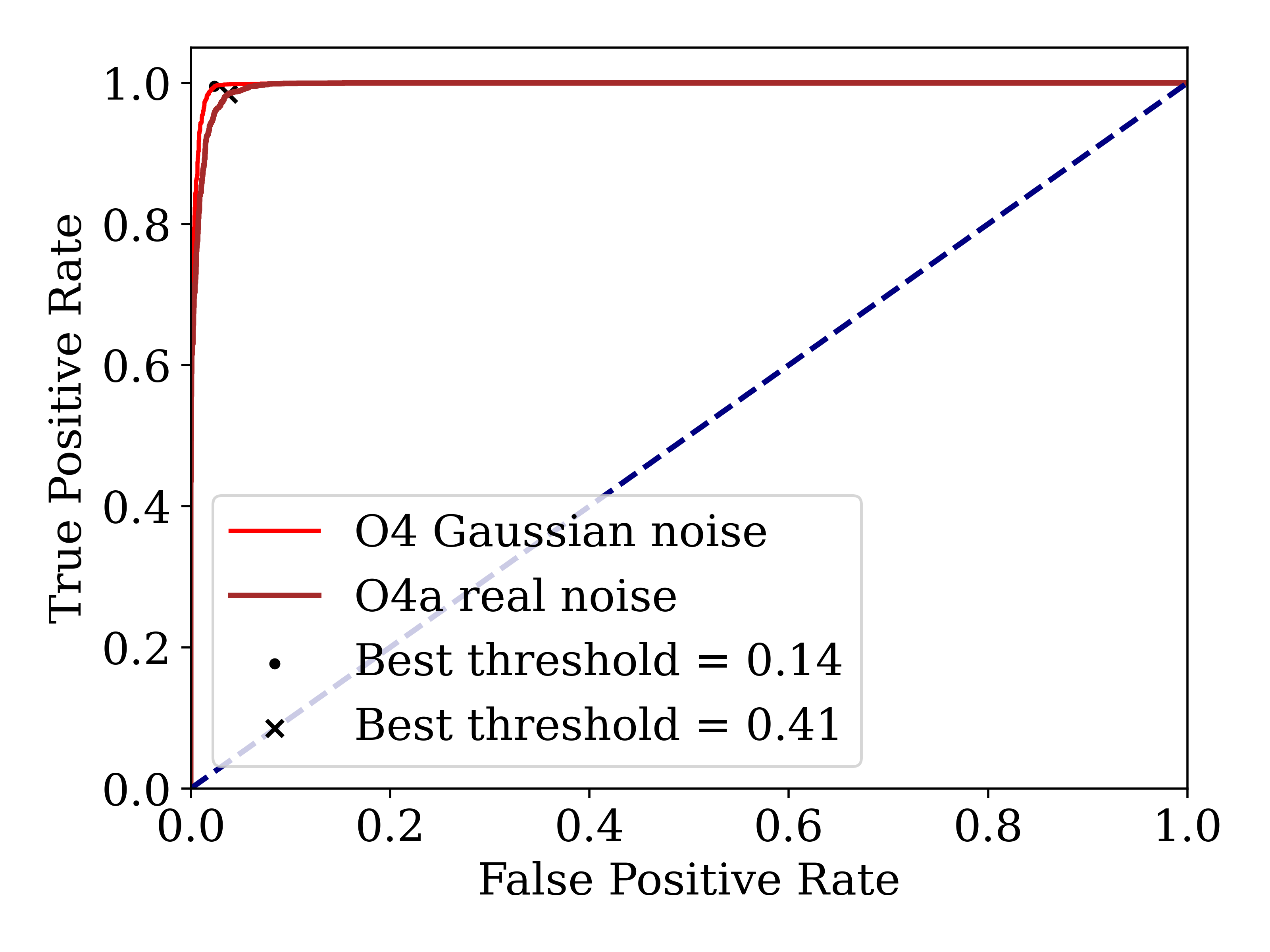}
        \caption{ROC curves for the classification of microlensed non-spinning (ML) and unlensed precessing (UP) gravitational-wave signals in O4 Gaussian noise (red) and O4a real detector noise (brown). Markers indicate the optimal decision thresholds selected for each dataset, with the corresponding threshold values annotated on the figure.
}
        \label{fig:3.1}
\end{figure}

CNN-2 was trained on QT images with signals injected in O4 Gaussian and O4a real noise. Figure \ref{fig:3.1} shows the ROC curves for the classification of microlensed non-spinning (ML) and unlensed precessing (UP) signals using CNN-2 trained on QT images generated in O4 Gaussian noise and O4a real detector noise. The results are obtained on a simulated test set consisting of 8000 signals (4000 per class). The classifier performs well in both cases, with ROC curves lying close to the upper-left region of the plot, indicating good separation between the two classes over a range of thresholds. The optimal classification thresholds are 0.14 for the Gaussian-noise model and 0.41 for the real-noise model. The model trained on real detector noise shows a slight reduction in performance at very low false-positive rates compared to the Gaussian-noise case, while the overall ROC behaviour remains similar between the two.

It should be noted that evaluation on real gravitational-wave events is currently limited by the lack of a sufficiently large sample of high-SNR unlensed precessing signals, which restricts direct validation beyond simulated data.

\begin{figure}
        \includegraphics[width=\columnwidth]{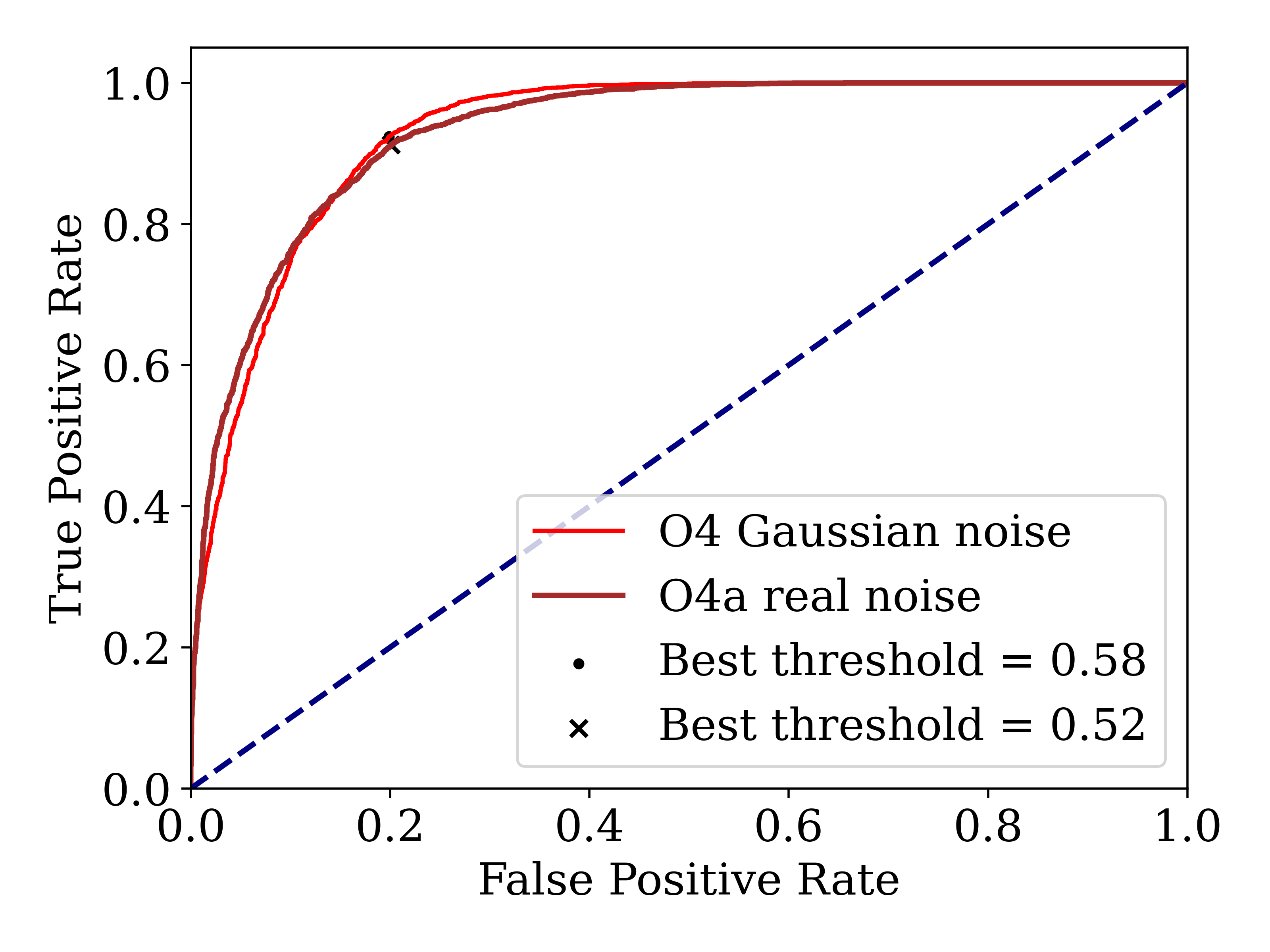}
        \caption{ROC curves for the classification of microlensed non-spinning (ML) and unlensed non-spinning (UN) gravitational-wave signals in O4 Gaussian noise (red) and O4a real detector noise (brown). Markers indicate the optimal decision thresholds selected for each dataset, with the corresponding threshold values annotated on the figure.
}
        \label{fig:3.2}
\end{figure}

\subsection{Microlensed non-spinning vs. Unlensed non-spinning}
Given the poor performance of CNN-1 and its inability to extract meaningful features from images of FS injected in noise, this representation was excluded from all subsequent experiments. As before, we report the performance of CNN-2 trained on QT images in Gaussian and real noise. Figure \ref{fig:3.2} shows the ROC curves for the classification of microlensed non-spinning (ML) and unlensed non-spinning (UN) signals using CNN-2 trained on QT images generated in O4 Gaussian noise and O4a real detector noise. The classifier exhibits strong performance in both cases, with the ROC curves lying close to the upper-left corner of the plot and yielding nearly identical classification accuracy. The optimal decision thresholds are 0.58 for the Gaussian-noise model and 0.52 for the real-noise model. While the Gaussian-noise model performs marginally better at intermediate false-positive rates, the close agreement between the two ROC curves indicates that the network remains effective at distinguishing ML and UN signals even in the presence of realistic detector noise. This demonstrates that the lensing-induced features learned from QT representations are robust against non-Gaussian noise artifacts present in O4a data.

\begin{figure*}
        \includegraphics[width=\textwidth]{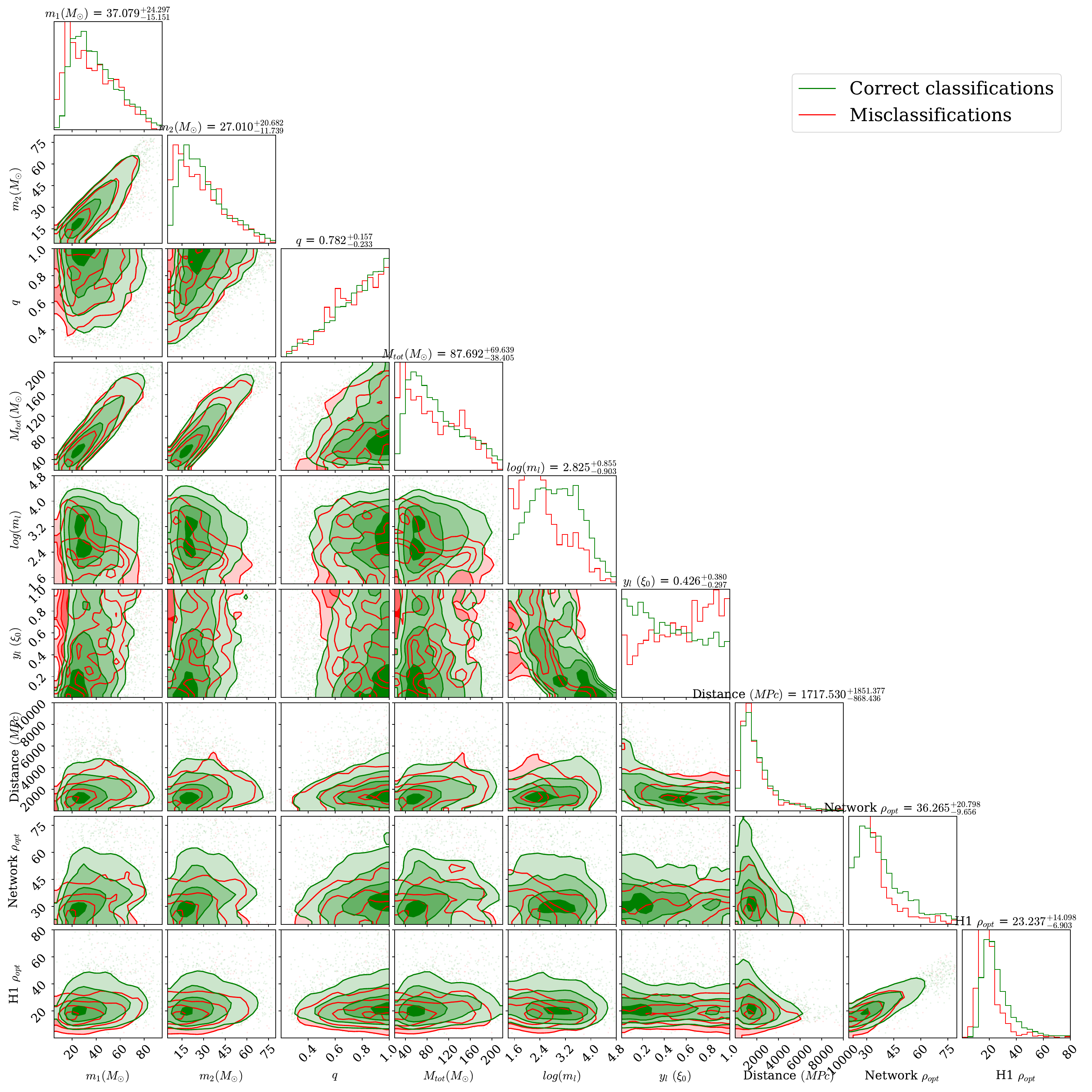}
        \caption{Corner plot showing the relations between different parameters for predictions on simulated test microlensed non-spinning (ML) signals injected in O4a real noise. It can be seen that misclassified ML signals tended to cluster in regions with high impact parameter and low lens mass.}
        \label{fig:3.3}
\end{figure*}

\begin{figure*}
        \includegraphics[width=\textwidth]{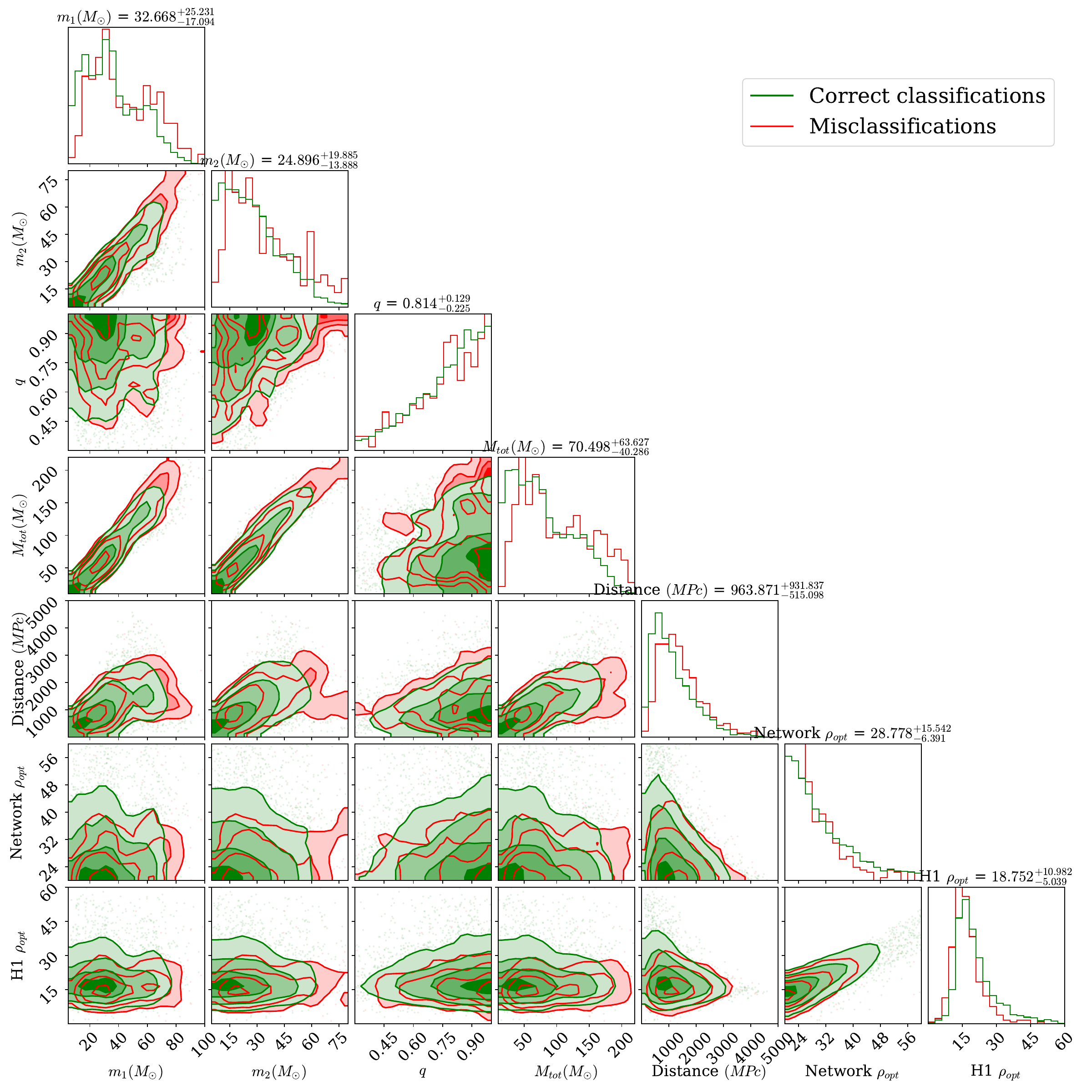}
        \caption{Corner plot showing the relations between different parameters for predictions on simulated test unlensed non-spinning (UN) signals injected in O4a real noise. It is seen that misclassified UN signals cluster in regions of high total mass and extreme mass ratios.}
        \label{fig:3.4}
\end{figure*}

To explore the origins of misclassifications, corner plot analyses were conducted using physical parameters of the test signals. Misclassified ML signals tended to cluster in regions with high impact parameter ($y_{\rm l}$) and low lens mass ($m_{\rm l}$) as indicated by Fig. \ref{fig:3.3}, where microlensing effects are milder and thus harder to detect. On the other hand, misclassified UN signals were more common in regions of high total mass ($M_{\rm {tot}}$) and extreme mass ratios ($q$) as inferred from Fig. \ref{fig:3.4}, where the waveform morphology could resemble that of microlensed signals. These scenarios point to parameter-space regions where classification is inherently more challenging.

\begin{figure}
        \includegraphics[scale=0.38]{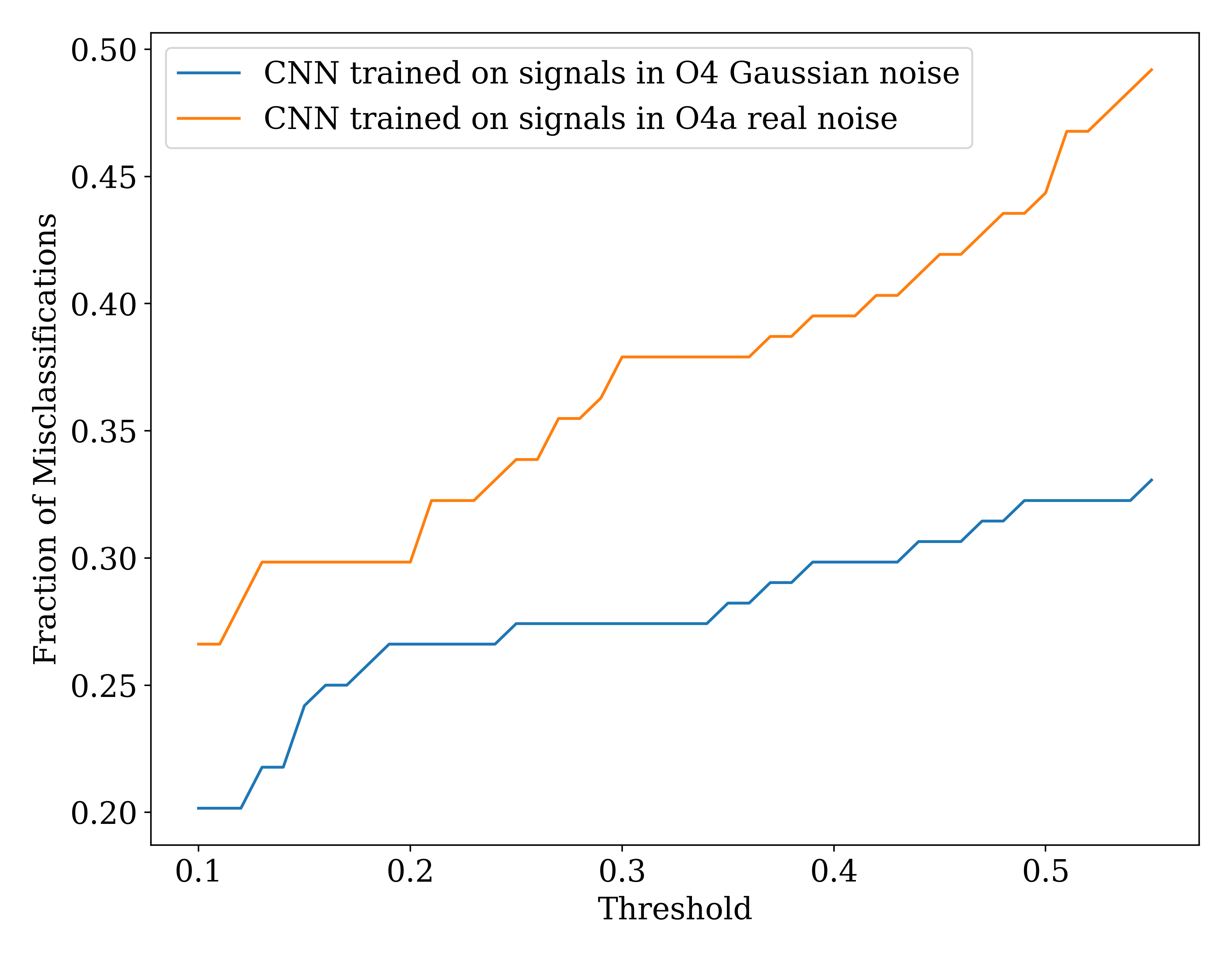}
        \caption{Fraction of misclassifications as a function CNN threshold for 124 real GW events observed by the H1 detector during the O4a observing run, using CNNs trained on QTs injected in O4 Gaussian noise (blue) and O4a real noise (orange). The CNN output represents the predicted probability of the unlensed class (class 1). Events are classified as unlensed when the predicted probability exceeds the threshold. Since all test samples are unlensed, the misclassification fraction increases with increasing threshold.}
        \label{fig:3.5}
\end{figure}

To evaluate the robustness of the trained models, their performance was tested on 124 real GW events from the O4a observing run of the LIGO-Virgo-Kagra collaboration. We restricted the evaluation to events observed with H1 detector.  Since microlensing of GWs has not yet been observed, the majority of events are expected to be classified as UN. The model trained on QTs injected in Gaussian noise performed reasonably, classifying 84 events as UN and 40 as ML at its optimal threshold of 0.58. In contrast, the model trained on QTs injected in real detector noise classifies 66 events as UN and 58 as ML at its optimal threshold of 0.52, indicating a substantially reduced discriminative capability. The fraction of misclassified events, defined as true UN events assigned to the ML class (0), is shown as a function of the decision threshold in Fig.~\ref{fig:3.5} (see Section~\ref{sec:2.2} for the probability output and thresholding convention). As expected from this definition, this fraction increases with increasing threshold, reflecting the progressively stricter criterion for assigning the UN class. Nevertheless, the model trained on O4a real-noise retains a substantially larger fraction of ML classifications over the threshold range, demonstrating poorer discrimination between the two classes. Taken together, these results suggest that the model trained on QTs injected in real noise lacks robustness and struggles to generalize beyond the training data. Further improvements to the model architecture or training procedure may be necessary to enhance its reliability in real-world applications. It should be noted, however, that these conclusions are drawn from a relatively small sample of events and are therefore subject to larger statistical uncertainties than our test dataset, which contain thousands of simulated signals. In the future, the threshold to be adopted for the analysis of real GW events will be determined through a dedicated large-scale background study, involving $\mathcal{O}(10^3)$ unlensed events, with comparable detector sensitivity and noise characteristics. Such a study is required to robustly calibrate the expected false-positive rate and select an operating threshold appropriate for real data analyses.

\section{Summary and Conclusions}
\label{sec:sum}
Microlensing and other GW signals with complex physics such as eccentricity and precession may cause similar modulations.  In this work, we particularly explored degeneracies of microlensed signals with precessing binaries.
Spin-induced precession in BBH systems can help enhance the precision of distance estimation during parameter inference. Additionally, it provides valuable information about the formation pathways of compact binaries and offers a means to test the fundamental properties of black holes. On the other hand, studying microlensed BBH signals allows us to probe the population and mass distribution of compact objects---such as primordial black holes or dense stellar remnants---along the line of sight. If unaccounted for, microlensing-induced waveform distortions may be misinterpreted as deviations from general relativity or introduce systematic biases in parameter inference, making accurate classification essential.

Here, we attempt to develop neural networks in order to distinguish the microlensed (ML) signals from unlensed precessing (UP) as well as unlensed non-spinning (UN) GW signals since the latter class is more abundantly found in the observations. To this end, we simulate 20,000 GW signals each belonging to three classes $-$ UN, UP and ML (without precession. Physically motivated prior distributions are used to generate the parameters for the signals, restricting the network optimal SNR~$>20$ and microlensing time delays, below 0.15~s. Two convolutional neural networks are designed to train on the input images showing the signals in the Frequency Series (FS) and Q-Transform (QT) format.

The classification of ML vs. UP is excellent in the FS representation only for the noise-free scenario --- the presence of noise led the model to confusion, with no meaningful patterns learned. The remainder of the analyses is then carried out with only QTs as input representations. The network performance on signals injected in O4 Gaussian noise is very promising for the classifications of ML vs. UP and UN, with accuracies reaching 95\%. However, on the signals injected in real detector noise, the network performance reached only a decent accuracy of 80\% for both these classifications.

We investigate the misclassifications by analysing the BBH and lensing parameter space. On the test data, we find that the majority of the ML signals with high impact parameter and low lens mass are falsely predicted as UN signals. This region corresponds to mild microlensing effects where it is possible for ML signals to be confused with UN signals. Moreover, most of the misclassified UN signals were located in a part of the parameter space where both the mass ratio and total mass of the binary was high. Again, this is the region where the chirps are short lived and can resemble microlensing induced modulations.

To further evaluate the effectiveness of the ML vs. UN network, we test on the sample of 124 real GW events from the O4a observing run closely resembling the training dataset. We find that the model trained on signals in O4 Gaussian noise, classifies most of them correctly as UN, but the model trained on signals in O4a real noise, does not perform as well.
For completeness, we also carry out the classification of UN vs. UP, but find that the performance is poor even for signals in O4 Gaussian noise. Improvements in neural network architectures and/or input representations are required for more accurate classification since the imprints of precession on the GW signals are not prominent. We note that the analysis is performed for GW signals whose network optimal SNR~$>20$. At lower SNRs, the data quality and visibility of QTs degrades, diminishing the model's accuracy, where using alternative data representations, de-noising or pre-processing may be required to improve performance.

There have been previous efforts to distinguish UN signals from ML signals using machine learning \citep[]{Kim_2021}, and to distinguish precession effects from microlensing using statistical methods \citep[]{kim2023discern}. But they do not account for the wave-optics effects in microlensing, which is very crucial at LIGO frequencies and are restricted to lens mass $10^2 - 10^5 M_\odot$, while also using only Gaussian noise conditions. In our knowledge, this work is the first study to incorporate wave-optics effects in simulating microlensed signals for the purpose of exploring the degeneracy between microlensing and precession, while also employing a statistically realistic populations and testing performance on real detector noises.

Our models already demonstrate promising performance in identifying key features relevant to distinguishing signals influenced by microlensing and precession. In future work, we aim to further enhance the accuracy and efficiency of our networks, and extend the analysis to more general scenarios --- specifically, testing the feasibility of distinguishing circular BBH signals from those affected by microlensing, precession, or both.

\begin{acknowledgments}
We thank Anuj Mishra for useful advice and suggestions regarding generation of GW signals and extraction of real detector noise segments. We also thank Sourabh Magare for helpful discussions on the machine learning aspects of the training framework. We acknowledge the use of Sarathi Cluster, which is part of the LIGO scientific collaboration data grid facility, for all the computing work done in this paper.
This material is based upon work supported by NSF's LIGO Laboratory which is a major facility fully funded by the National Science Foundation.
\end{acknowledgments}

\section*{Data Availability}

The data generated and analyzed in this study will be made available from the corresponding author upon reasonable request.

\bibliographystyle{apsrev4-2}
\bibliography{paper}

@misc{buonanno2015sourcesgravitationalwavestheory,
      title={Sources of Gravitational Waves: Theory and Observations}, 
      author={Alessandra Buonanno and B. S. Sathyaprakash},
      year={2015},
      eprint={1410.7832},
      archivePrefix={arXiv},
      primaryClass={gr-qc},
      url={https://arxiv.org/abs/1410.7832}, 
}

@article{PhysRevLett.116.061102,
  title = {Observation of Gravitational Waves from a Binary Black Hole Merger},
  author = {Abbott, B. P. and others},
  collaboration = {LIGO Scientific Collaboration and Virgo Collaboration},
  journal = {Phys. Rev. Lett.},
  volume = {116},
  issue = {6},
  pages = {061102},
  numpages = {16},
  year = {2016},
  month = {Feb},
  publisher = {American Physical Society},
  doi = {10.1103/PhysRevLett.116.061102},
  url = {https://link.aps.org/doi/10.1103/PhysRevLett.116.061102}
}

@article{PhysRevX.13.041039,
  title = {GWTC-3: Compact Binary Coalescences Observed by LIGO and Virgo during the Second Part of the Third Observing Run},
  author = {Abbott, R. and others},
  collaboration = {LIGO Scientific Collaboration, Virgo Collaboration, and KAGRA Collaboration},
  journal = {Phys. Rev. X},
  volume = {13},
  issue = {4},
  pages = {041039},
  numpages = {89},
  year = {2023},
  month = {Dec},
  publisher = {American Physical Society},
  doi = {10.1103/PhysRevX.13.041039},
  url = {https://link.aps.org/doi/10.1103/PhysRevX.13.041039}
}

@article{Liu_2024,
   title={Can we discern millilensed gravitational-wave signals from signals produced by precessing binary black holes with ground-based detectors?},
   volume={110},
   ISSN={2470-0029},
   url={http://dx.doi.org/10.1103/PhysRevD.110.123008},
   DOI={10.1103/physrevd.110.123008},
   number={12},
   journal={Physical Review D},
   publisher={American Physical Society (APS)},
   author={Liu, Anna and Kim, Kyungmin},
   year={2024},
   month=dec }

@article{PhysRevD.49.6274,
  title = {Spin-induced orbital precession and its modulation of the gravitational waveforms from merging binaries},
  author = {Apostolatos, Theocharis A. and Cutler, Curt and Sussman, Gerald J. and Thorne, Kip S.},
  journal = {Phys. Rev. D},
  volume = {49},
  issue = {12},
  pages = {6274--6297},
  numpages = {0},
  year = {1994},
  month = {Jun},
  publisher = {American Physical Society},
  doi = {10.1103/PhysRevD.49.6274},
  url = {https://link.aps.org/doi/10.1103/PhysRevD.49.6274}
}

@article{PhysRevD.110.023038,
  title = {Searching for gravitational-wave signals from precessing black hole binaries with the GstLAL pipeline},
  author = {Schmidt, Stefano and Caudill, Sarah and Creighton, Jolien D. E. and Magee, Ryan and Tsukada, Leo and others},
  journal = {Phys. Rev. D},
  volume = {110},
  issue = {2},
  pages = {023038},
  numpages = {24},
  year = {2024},
  month = {Jul},
  publisher = {American Physical Society},
  doi = {10.1103/PhysRevD.110.023038},
  url = {https://link.aps.org/doi/10.1103/PhysRevD.110.023038}
}

@article{Meena_2022,
   title={Gravitational lensing of gravitational waves: Probability of microlensing in galaxy-scale lens population},
   volume={517},
   ISSN={1365-2966},
   url={http://dx.doi.org/10.1093/mnras/stac2721},
   DOI={10.1093/mnras/stac2721},
   number={1},
   journal={Monthly Notices of the Royal Astronomical Society},
   publisher={Oxford University Press (OUP)},
   author={Meena, Ashish Kumar and Mishra, Anuj and More, Anupreeta and Bose, Sukanta and Bagla, Jasjeet Singh},
   year={2022},
   month=sep, pages={872–884} }

@article{meena2019,
    author = {Meena, Ashish Kumar and Bagla, Jasjeet Singh},
    title = {Gravitational lensing of gravitational waves: wave nature and prospects for detection},
    journal = {Monthly Notices of the Royal Astronomical Society},
    volume = {492},
    number = {1},
    pages = {1127-1134},
    year = {2019},
    month = {12},
    issn = {0035-8711},
    doi = {10.1093/mnras/stz3509},
    url = {https://doi.org/10.1093/mnras/stz3509},
    eprint = {https://academic.oup.com/mnras/article-pdf/492/1/1127/31776880/stz3509.pdf},
}

@misc{kim2023discern,
      title={Can we discern microlensed gravitational-wave signals from the signal of precessing compact binary mergers?}, 
      author={Kyungmin Kim and Anna Liu},
      year={2023},
      eprint={2301.07253},
      archivePrefix={arXiv},
      primaryClass={gr-qc}
}

@article{Cuoco_2020,
   title={Enhancing gravitational-wave science with machine learning},
   volume={2},
   ISSN={2632-2153},
   url={http://dx.doi.org/10.1088/2632-2153/abb93a},
   DOI={10.1088/2632-2153/abb93a},
   number={1},
   journal={Machine Learning: Science and Technology},
   publisher={IOP Publishing},
   author={Cuoco, Elena and Powell, Jade and Cavaglià, Marco and Ackley, Kendall and Bejger, Michał and Chatterjee, Chayan and Coughlin, Michael and Coughlin, Scott and Easter, Paul and Essick, Reed and Gabbard, Hunter and Gebhard, Timothy and Ghosh, Shaon and Haegel, Leïla and Iess, Alberto and Keitel, David and Márka, Zsuzsa and Márka, Szabolcs and Morawski, Filip and Nguyen, Tri and Ormiston, Rich and Pürrer, Michael and Razzano, Massimiliano and Staats, Kai and Vajente, Gabriele and Williams, Daniel},
   year={2020},
   month=dec, pages={011002} }

@article{PhysRevD.103.104056,
  title = {Computationally efficient models for the dominant and subdominant harmonic modes of precessing binary black holes},
  author = {Pratten, Geraint and Garc\'{\i}a-Quir\'os, Cecilio and Colleoni, Marta and Ramos-Buades, Antoni and Estell\'es, H\'ector and Mateu-Lucena, Maite and Jaume, Rafel and Haney, Maria and Keitel, David and Thompson, Jonathan E. and Husa, Sascha},
  journal = {Phys. Rev. D},
  volume = {103},
  issue = {10},
  pages = {104056},
  numpages = {36},
  year = {2021},
  month = {May},
  publisher = {American Physical Society},
  doi = {10.1103/PhysRevD.103.104056},
  url = {https://link.aps.org/doi/10.1103/PhysRevD.103.104056}
}

@ARTICLE{2016LRR....19....1A,
       author = {{Abbott}, B.~P. and {LIGO Scientific Collaboration} and {Virgo Collaboration}},
        title = "{Prospects for Observing and Localizing Gravitational-Wave Transients with Advanced LIGO and Advanced Virgo}",
      journal = {Living Reviews in Relativity},
         year = 2016,
        month = dec,
       volume = {19},
       number = {1},
          eid = {1},
        pages = {1},
          doi = {10.1007/lrr-2016-1},
       adsurl = {https://ui.adsabs.harvard.edu/abs/2016LRR....19....1A}
}

@article{PhysRevD.98.083005,
  title = {Discovering intermediate-mass black hole lenses through gravitational wave lensing},
  author = {Lai, Kwun-Hang and Hannuksela, Otto A. and Herrera-Mart\'{\i}n, Antonio and Diego, Jose M. and Broadhurst, Tom and Li, Tjonnie G. F.},
  journal = {Phys. Rev. D},
  volume = {98},
  issue = {8},
  pages = {083005},
  numpages = {7},
  year = {2018},
  month = {Oct},
  publisher = {American Physical Society},
  doi = {10.1103/PhysRevD.98.083005},
  url = {https://link.aps.org/doi/10.1103/PhysRevD.98.083005}
}

@article{annurev:/content/journals/10.1146/annurev-astro-081811-125615,
   author = "Madau, Piero and Dickinson, Mark",
   title = "Cosmic Star-Formation History", 
   journal= "Annual Review of Astronomy and Astrophysics",
   year = "2014",
   volume = "52",
   number = "Volume 52, 2014",
   pages = "415-486",
   doi = "https://doi.org/10.1146/annurev-astro-081811-125615",
   url = "https://www.annualreviews.org/content/journals/10.1146/annurev-astro-081811-125615",
   publisher = "Annual Reviews",
   issn = "1545-4282",
   type = "Journal Article",
  }

@ARTICLE{2018ApJ...863L..41F,
       author = {{Fishbach}, Maya and {Holz}, Daniel E. and {Farr}, Will M.},
        title = "{Does the Black Hole Merger Rate Evolve with Redshift?}",
      journal = {\apjl},
         year = 2018,
        month = aug,
       volume = {863},
       number = {2},
          eid = {L41},
        pages = {L41},
          doi = {10.3847/2041-8213/aad800},
archivePrefix = {arXiv},
       eprint = {1805.10270},
 primaryClass = {astro-ph.HE},
       adsurl = {https://ui.adsabs.harvard.edu/abs/2018ApJ...863L..41F}
}

@article{bilby_paper,
    author = "Ashton, Gregory and others",
    title = "{BILBY: A user-friendly Bayesian inference library for gravitational-wave astronomy}",
    eprint = "1811.02042",
    archivePrefix = "arXiv",
    primaryClass = "astro-ph.IM",
    doi = "10.3847/1538-4365/ab06fc",
    journal = "Astrophys. J. Suppl.",
    volume = "241",
    number = "2",
    pages = "27",
    year = "2019"
}

@ARTICLE{2019PhRvD.100d3030T,
  author = {{Talbot}, Colm and {Smith}, Rory and {Thrane}, Eric and {Poole}, Gregory B.},
  title = "{Parallelized inference for gravitational-wave astronomy}",
  journal = {\prd},
  year = 2019,
  month = aug,
  volume = {100},
  number = {4},
  eid = {043030},
  pages = {043030},
  doi = {10.1103/PhysRevD.100.043030},
  archivePrefix = {arXiv},
  eprint = {1904.02863},
  primaryClass = {astro-ph.IM},
}

@article{Magare_2024,
   title={SLICK: Strong Lensing Identification of Candidates Kindred in gravitational wave data},
   volume={535},
   ISSN={1365-2966},
   url={http://dx.doi.org/10.1093/mnras/stae2408},
   DOI={10.1093/mnras/stae2408},
   number={1},
   journal={Monthly Notices of the Royal Astronomical Society},
   publisher={Oxford University Press (OUP)},
   author={Magare, Sourabh and More, Anupreeta and Choudhary, Sunil},
   year={2024},
   month=oct, pages={990–999} }

@misc{kingma2017adammethodstochasticoptimization,
      title={Adam: A Method for Stochastic Optimization}, 
      author={Diederik P. Kingma and Jimmy Ba},
      year={2017},
      eprint={1412.6980},
      archivePrefix={arXiv},
      primaryClass={cs.LG},
      url={https://arxiv.org/abs/1412.6980}, 
}

@misc{chollet2015keras,
  title={Keras},
  author={Chollet, Fran\c{c}ois and others},
  year={2015},
  howpublished={\url{https://keras.io}},
}

@article{scikit-learn,
  title={Scikit-learn: Machine Learning in {P}ython},
  author={Pedregosa, F. and Varoquaux, G. and Gramfort, A. and Michel, V.
          and Thirion, B. and Grisel, O. and Blondel, M. and Prettenhofer, P.
          and Weiss, R. and Dubourg, V. and Vanderplas, J. and Passos, A. and
          Cournapeau, D. and Brucher, M. and Perrot, M. and Duchesnay, E.},
  journal={Journal of Machine Learning Research},
  volume={12},
  pages={2825--2830},
  year={2011}
}

@article{Kim_2021,
   title={Identification of Lensed Gravitational Waves with Deep Learning},
   volume={915},
   ISSN={1538-4357},
   url={http://dx.doi.org/10.3847/1538-4357/ac0143},
   DOI={10.3847/1538-4357/ac0143},
   number={2},
   journal={The Astrophysical Journal},
   publisher={American Astronomical Society},
   author={Kim, Kyungmin and Lee, Joongoo and Yuen, Robin S. H. and Hannuksela, Otto A. and Li, Tjonnie G. F.},
   year={2021},
   month=jul, pages={119} }

@misc{cuoco2024,
      title={Applications of machine learning in gravitational wave research with current interferometric detectors}, 
      author={Elena Cuoco and Marco Cavaglià and Ik Siong Heng and David Keitel and Christopher Messenger},
      year={2024},
      eprint={2412.15046},
      archivePrefix={arXiv},
      primaryClass={gr-qc},
      url={https://arxiv.org/abs/2412.15046}, 
}

@article{PhysRevD.104.124057,
  title = {Rapid identification of strongly lensed gravitational-wave events with machine learning},
  author = {Goyal, Srashti and D., Harikrishnan and Kapadia, Shasvath J. and Ajith, Parameswaran},
  journal = {Phys. Rev. D},
  volume = {104},
  issue = {12},
  pages = {124057},
  numpages = {12},
  year = {2021},
  month = {Dec},
  publisher = {American Physical Society},
  doi = {10.1103/PhysRevD.104.124057},
  url = {https://link.aps.org/doi/10.1103/PhysRevD.104.124057}
}

@article{PhysRevX.13.011048,
  title = {Population of Merging Compact Binaries Inferred Using Gravitational Waves through GWTC-3},
  author = {Abbott, R. and others},
  collaboration = {LIGO Scientific Collaboration, Virgo Collaboration, and KAGRA Collaboration},
  journal = {Phys. Rev. X},
  volume = {13},
  issue = {1},
  pages = {011048},
  numpages = {75},
  year = {2023},
  month = {Mar},
  publisher = {American Physical Society},
  doi = {10.1103/PhysRevX.13.011048},
  url = {https://link.aps.org/doi/10.1103/PhysRevX.13.011048}
}

@article{PhysRevD.107.024030,
  title = {Deep learning network to distinguish binary black hole signals from short-duration noise transients},
  author = {Choudhary, Sunil and More, Anupreeta and Suyamprakasam, Sudhagar and Bose, Sukanta},
  journal = {Phys. Rev. D},
  volume = {107},
  issue = {2},
  pages = {024030},
  numpages = {12},
  year = {2023},
  month = {Jan},
  publisher = {American Physical Society},
  doi = {10.1103/PhysRevD.107.024030},
  url = {https://link.aps.org/doi/10.1103/PhysRevD.107.024030}
}

@ARTICLE{2026arXiv260527225T,
       author = {{The LIGO Scientific Collaboration} and {the Virgo Collaboration} and {the KAGRA Collaboration}},
        title = "{GWTC-5.0: Observations from the Second Part of the Fourth LIGO-Virgo-KAGRA Observing Run and Updates to the Gravitational-Wave Transient Catalog}",
      journal = {arXiv e-prints},
         year = 2026,
        month = may,
          eid = {arXiv:2605.27225},
        pages = {arXiv:2605.27225},
          doi = {10.48550/arXiv.2605.27225},
archivePrefix = {arXiv},
       eprint = {2605.27225},
 primaryClass = {gr-qc},
       adsurl = {https://ui.adsabs.harvard.edu/abs/2026arXiv260527225T}
}

@ARTICLE{2025arXiv250818082T,
       author = {{The LIGO Scientific Collaboration} and {the Virgo Collaboration} and {the KAGRA Collaboration}},
        title = "{GWTC-4.0: Updating the Gravitational-Wave Transient Catalog with Observations from the First Part of the Fourth LIGO-Virgo-KAGRA Observing Run}",
      journal = {arXiv e-prints},
         year = 2025,
        month = aug,
          eid = {arXiv:2508.18082},
        pages = {arXiv:2508.18082},
          doi = {10.48550/arXiv.2508.18082},
archivePrefix = {arXiv},
       eprint = {2508.18082},
 primaryClass = {gr-qc},
       adsurl = {https://ui.adsabs.harvard.edu/abs/2025arXiv250818082T}
}

@article{Hannam_2022,
   title={General-relativistic precession in a black-hole binary},
   volume={610},
   ISSN={1476-4687},
   url={http://dx.doi.org/10.1038/s41586-022-05212-z},
   DOI={10.1038/s41586-022-05212-z},
   number={7933},
   journal={Nature},
   publisher={Springer Science and Business Media LLC},
   author={Hannam, Mark and Hoy, Charlie and Thompson, Jonathan E. and Fairhurst, Stephen and Raymond, Vivien and Colleoni, Marta and Davis, Derek and Estellés, Héctor and Haster, Carl-Johan and Helmling-Cornell, Adrian and Husa, Sascha and Keitel, David and Massinger, T. J. and Menéndez-Vázquez, Alexis and Mogushi, Kentaro and Ossokine, Serguei and Payne, Ethan and Pratten, Geraint and Romero-Shaw, Isobel and Sadiq, Jam and Schmidt, Patricia and Tenorio, Rodrigo and Udall, Richard and Veitch, John and Williams, Daniel and Yelikar, Anjali Balasaheb and Zimmerman, Aaron},
   year={2022},
   month=Oct, pages={652–655} }

@article{PhysRevD.102.043015,
  title = {GW190412: Observation of a binary-black-hole coalescence with asymmetric masses},
  author = {Abbott, R. and others},
  collaboration = {LIGO Scientific Collaboration and Virgo Collaboration},
  journal = {Phys. Rev. D},
  volume = {102},
  issue = {4},
  pages = {043015},
  numpages = {29},
  year = {2020},
  month = {Aug},
  publisher = {American Physical Society},
  doi = {10.1103/PhysRevD.102.043015},
  url = {https://link.aps.org/doi/10.1103/PhysRevD.102.043015}
}

@ARTICLE{2003PhRvD..67j4025B,
       author = {{Buonanno}, Alessandra and {Chen}, Yanbei and {Vallisneri}, Michele},
        title = "{Detecting gravitational waves from precessing binaries of spinning compact objects: Adiabatic limit}",
      journal = {\prd},
         year = 2003,
        month = may,
       volume = {67},
       number = {10},
          eid = {104025},
        pages = {104025},
          doi = {10.1103/PhysRevD.67.104025},
archivePrefix = {arXiv},
       eprint = {gr-qc/0211087},
 primaryClass = {gr-qc},
       adsurl = {https://ui.adsabs.harvard.edu/abs/2003PhRvD..67j4025B}
}

@article{wright2022gravelamps,
  title = {Gravelamps: Gravitational Wave Lensing Mass Profile Model Selection},
  author = {Mick Wright and Martin Hendry},
  year = {2022},
  publisher = {American Astronomical Society},
  journal = {The Astrophysical Journal},
  volume = {935},
  pages = {68},
  doi = {10.3847/1538-4357/ac7ec2},
  url = {https://doi.org/10.3847/1538-4357/ac7ec2}
}

@ARTICLE{2021ApJ...923...14A,
       author = {{Abbott}, R. and others},
        title = "{Search for Lensing Signatures in the Gravitational-Wave Observations from the First Half of LIGO─Virgo's Third Observing Run}",
      journal = {\apj},
         year = 2021,
        month = dec,
       volume = {923},
       number = {1},
          eid = {14},
        pages = {14},
          doi = {10.3847/1538-4357/ac23db},
archivePrefix = {arXiv},
       eprint = {2105.06384},
 primaryClass = {gr-qc},
       adsurl = {https://ui.adsabs.harvard.edu/abs/2021ApJ...923...14A}
}

@ARTICLE{2024ApJ...970..191A,
       author = {{Abbott}, R. and others},
        title = "{Search for Gravitational-lensing Signatures in the Full Third Observing Run of the LIGO─Virgo Network}",
      journal = {\apj},
         year = 2024,
        month = aug,
       volume = {970},
       number = {2},
          eid = {191},
        pages = {191},
          doi = {10.3847/1538-4357/ad3e83},
archivePrefix = {arXiv},
       eprint = {2304.08393},
 primaryClass = {gr-qc},
       adsurl = {https://ui.adsabs.harvard.edu/abs/2024ApJ...970..191A}
}

@ARTICLE{2023MNRAS.526.3832J,
       author = {{Janquart}, J. and {Wright}, M. and {Goyal}, S. and {Chan}, J.~C.~L. and {Ganguly}, A. and {Garr{\'o}n}, {\'A}. and {Keitel}, D. and {Li}, A.~K.~Y. and {Liu}, A. and {Lo}, R.~K.~L. and {Mishra}, A. and {More}, A. and {Phurailatpam}, H. and {Prasia}, P. and {Ajith}, P. and {Biscoveanu}, S. and {Cremonese}, P. and {Cudell}, J.~R. and {Ezquiaga}, J.~M. and {Garcia-Bellido}, J. and {Hannuksela}, O.~A. and {Haris}, K. and {Harry}, I. and {Hendry}, M. and {Husa}, S. and {Kapadia}, S. and {Li}, T.~G.~F. and {Maga{\~n}a Hernandez}, I. and {Mukherjee}, S. and {Seo}, E. and {Van Den Broeck}, C. and {Veitch}, J.},
        title = "{Follow-up analyses to the O3 LIGO-Virgo-KAGRA lensing searches}",
      journal = {\mnras},
         year = 2023,
        month = dec,
       volume = {526},
       number = {3},
        pages = {3832-3860},
          doi = {10.1093/mnras/stad2909},
archivePrefix = {arXiv},
       eprint = {2306.03827},
 primaryClass = {gr-qc},
       adsurl = {https://ui.adsabs.harvard.edu/abs/2023MNRAS.526.3832J}
}

@ARTICLE{2025arXiv251216347T,
       author = {{The LIGO Scientific Collaboration} and {the Virgo Collaboration} and {the KAGRA Collaboration}},
        title = "{GWTC-4.0: Searches for Gravitational-Wave Lensing Signatures}",
      journal = {arXiv e-prints},
         year = 2025,
        month = dec,
          eid = {arXiv:2512.16347},
        pages = {arXiv:2512.16347},
          doi = {10.48550/arXiv.2512.16347},
archivePrefix = {arXiv},
       eprint = {2512.16347},
 primaryClass = {gr-qc},
       adsurl = {https://ui.adsabs.harvard.edu/abs/2025arXiv251216347T}
}

@ARTICLE{2022ApJ...926L..28B,
       author = {{Basak}, S. and {Ganguly}, A. and {Haris}, K. and {Kapadia}, S. and {Mehta}, A.~K. and {Ajith}, P.},
        title = "{Constraints on Compact Dark Matter from Gravitational Wave Microlensing}",
      journal = {\apjl},
         year = 2022,
        month = feb,
       volume = {926},
       number = {2},
          eid = {L28},
        pages = {L28},
          doi = {10.3847/2041-8213/ac4dfa},
archivePrefix = {arXiv},
       eprint = {2109.06456},
 primaryClass = {gr-qc},
       adsurl = {https://ui.adsabs.harvard.edu/abs/2022ApJ...926L..28B}
}

@ARTICLE{2023PhRvD.108l3016M,
       author = {{McIsaac}, Connor and {Hoy}, Charlie and {Harry}, Ian},
        title = "{Search technique to observe precessing compact binary mergers in the advanced detector era}",
      journal = {\prd},
         year = 2023,
        month = dec,
       volume = {108},
       number = {12},
          eid = {123016},
        pages = {123016},
          doi = {10.1103/PhysRevD.108.123016},
archivePrefix = {arXiv},
       eprint = {2303.17364},
 primaryClass = {gr-qc},
       adsurl = {https://ui.adsabs.harvard.edu/abs/2023PhRvD.108l3016M}
}

@ARTICLE{2017PhRvD..95f4056I,
       author = {{Indik}, Nathaniel and {Haris}, K. and {Dal Canton}, Tito and {Fehrmann}, Henning and {Krishnan}, Badri and {Lundgren}, Andrew and {Nielsen}, Alex B. and {Pai}, Archana},
        title = "{Stochastic template bank for gravitational wave searches for precessing neutron-star-black-hole coalescence events}",
      journal = {\prd},
         year = 2017,
        month = mar,
       volume = {95},
       number = {6},
          eid = {064056},
        pages = {064056},
          doi = {10.1103/PhysRevD.95.064056},
archivePrefix = {arXiv},
       eprint = {1612.05173},
 primaryClass = {gr-qc},
       adsurl = {https://ui.adsabs.harvard.edu/abs/2017PhRvD..95f4056I}
}

@article{Chan_2025,
   title={Detectability of lensed gravitational waves in matched-filtering searches},
   volume={111},
   ISSN={2470-0029},
   url={http://dx.doi.org/10.1103/PhysRevD.111.084019},
   DOI={10.1103/physrevd.111.084019},
   number={8},
   journal={Physical Review D},
   publisher={American Physical Society (APS)},
   author={Chan, Juno C.L. and Seo, Eungwang and Li, Alvin K.Y. and Fong, Heather and Ezquiaga, Jose M.},
   year={2025},
   month=Apr }

@article{8cz1-kl6n,
  title = {Identification and characterization of distorted gravitational waves by lensing using deep learning},
  author = {Chan, Juno C. L. and Zertuche, Lorena Maga\~na and Ezquiaga, Jose Mar\'{\i}a and Lo, Rico K. L. and Vujeva, Luka and Bowman, Joey},
  journal = {Phys. Rev. D},
  volume = {113},
  issue = {2},
  pages = {024041},
  numpages = {17},
  year = {2026},
  month = {Jan},
  publisher = {American Physical Society},
  doi = {10.1103/8cz1-kl6n},
  url = {https://link.aps.org/doi/10.1103/8cz1-kl6n}
}

@ARTICLE{2025ApJ...984..107C,
       author = {{Chakraborty}, Aniruddha and {Mukherjee}, Suvodip},
        title = "{{\ensuremath{\mu}}-GLANCE: A Novel Technique to Detect Chromatically and Achromatically Lensed Gravitational-wave Signals}",
      journal = {\apj},
         year = 2025,
        month = may,
       volume = {984},
       number = {2},
          eid = {107},
        pages = {107},
          doi = {10.3847/1538-4357/adc578},
archivePrefix = {arXiv},
       eprint = {2410.06995},
 primaryClass = {gr-qc},
       adsurl = {https://ui.adsabs.harvard.edu/abs/2025ApJ...984..107C}
}

@ARTICLE{2025ApJ...990...68C,
       author = {{Chakraborty}, Aniruddha and {Mukherjee}, Suvodip},
        title = "{The First Model-independent Chromatic Microlensing Search: No Evidence in the Gravitational Wave Catalog of LIGO─Virgo─KAGRA}",
      journal = {\apj},
         year = 2025,
        month = sep,
       volume = {990},
       number = {1},
          eid = {68},
        pages = {68},
          doi = {10.3847/1538-4357/adf330},
archivePrefix = {arXiv},
       eprint = {2503.16281},
 primaryClass = {gr-qc},
       adsurl = {https://ui.adsabs.harvard.edu/abs/2025ApJ...990...68C}
}

\end{document}